\documentclass{jfm}
\usepackage{graphicx}
\usepackage{epstopdf, epsfig}
\usepackage{amsmath}
\usepackage{stackengine}
\usepackage{xcolor}
\usepackage{scalerel}
\usepackage{relsize}
\usepackage[export]{adjustbox}

\shorttitle{Bubble-induced EIT}
\shortauthor{Naseer et~al.}

\title{Bubbles-induced transition to elasto-inertial turbulence}

\author{Hafiz Usman Naseer\aff{1},
  Daulet Izbassarov\aff{2},
  Marco Edoardo Rosti\aff{3}
 \and Metin Muradoglu\aff{1}
 \corresp{\email{mmuradoglu@ku.edu.tr}}}

\affiliation{\aff{1}Department of Mechanical Engineering, Koç University,
34450 Istanbul, Türkiye
\aff{2}Finnish Meteorological Institute, Erik Palmenin aukio 1, 00560 Helsinki, Finland
\aff{3}Complex Fluids and Flows Unit, Okinawa Institute of Science and Technology Graduate University,
1919-1 Tancha, Onna-son, Okinawa 904-0495, Japan}

\begin{document}

\maketitle

\begin{abstract}

Interface-resolved direct numerical simulations are performed to investigate bubble-induced transition from laminar to elasto-inertial turbulent (EIT) state in a pressure-driven viscoelastic square channel flow. The Giesekus model is used to account for the viscoelasticity of the continuous phase while the dispersed phase is Newtonian. Simulations are performed for both single and two-phase flows for a wide range of the Reynolds ($Re$) and the Weissenberg ($Wi$) numbers. It is demonstrated that injection of bubbles into a laminar viscoelastic flow introduces streamline curvature that is sufficient to trigger an elastic instability leading to a transition to a fully EIT regime. The temporal turbulent kinetic energy spectrum shows a scaling of $-2$ for this multiphase EIT regime. It is shown that, once the flow is fully transitioned to a turbulent state by the injection of bubbles, the drag increases for all the cases. It is also observed that bubbles move towards the channel centreline and form a string-shaped alignment pattern in the core region at the lower values of $Re=10$ and $Wi=1$. In this regime, the flow exhibits an intermittent behaviour, i.e., there are turbulent like fluctuations in the core region while it is essentially laminar near the wall. Unlike the solid particles, it is found that increasing shear-thinning effect breaks up the alignment of bubbles. Interestingly, the drag remains slightly lower in this intermittent regime than the corresponding laminar state. 
 
\end{abstract}

\begin{keywords}
Elasto-inertial turbulence, elastic instability, viscoelasticity,  multiphase flows  
\end{keywords}

\section{Introduction}
Adding a minute amount of long-chain flexible polymers can drastically change mechanical properties of an otherwise Newtonian fluid and makes it viscoelastic. Stretching of polymer molecules causes elastic stresses and gives a memory to the fluid flow,  resulting in a range of exotic and often counter-intuitive behaviours such as rod climbing, elastic recoil and drag reduction in turbulent flows \citep{datta2022perspectives}. The exact mechanism behind many of the exotic phenomena exhibited by the viscoelastic fluids are yet to be fully explored. A striking example of such phenomena is sustaining a chaotic, turbulent like fluid motion observed at a low Reynolds number \citep{groisman2000elastic,steinberg2021elastic,datta2022perspectives, singh2024intermittency}. In case of a classical Newtonian turbulent flow, the flow instabilities arise from the non-linear convective terms of the Navier-Stokes equations, and therefore a very high Reynolds number is always required to sustain a turbulent flow field. In case of a viscoelastic fluid flow, however, the non-linear coupling between viscous and viscoelastic stresses are capable of triggering and sustaining a chaotic motion even in the absence of inertia~\citep{datta2022perspectives}. This turbulent flow state at a Reynolds number below the threshold value of inertial turbulence (IT) is often referred as the ``elasto-inertial turbulence" (EIT). When the Reynolds number is further reduced to a negligibly smaller value, it turns out that high enough viscoelastic stresses are still capable of sustaining a chaotic turbulent-like state, referred as the elastic turbulence (ET), which is purely dominated by the elastic effects (\citet{steinberg2021elastic}). The present study mainly focuses on the elasto-inertial instability and transition to the elasto-inertial turbulence where, as the name implies, the chaotic flow state is caused and sustained by both elastic and inertial effects.

In the inertialess limit, only nonlinearity arises from the elastic stress that becomes anisotropic in a shear flow and results in the hoop stress acting on the fluid in the direction of the streamline curvature. This hoop stress provides the driving force for the elastic instability and triggers a transition to the elastic turbulent (ET) state even in the absence of inertia~\citep{steinberg2021elastic}. Since the discovery of ET in inertialess flows with curved streamlines by \cite{groisman2000elastic}, a non-linear instability leading to a similar kind of chaotic state in planar shear flows was suspected and investigated by \cite{morozov2005subcritical} and \citet{morozov2019subcritical}. These studies suggested that a non-linear instability can also be triggered in a planar shear flow when a sufficient flow curvature is induced via external perturbations. To realize this, the initial perturbations have been usually induced in the form of blowing/suction at the wall or by placing an obstacle in the channel to trigger the instability and attain an EIT state (\cite{dubief2023elasto}).

There are different pathways which can be followed to achieve the EIT state in a wall bounded flow. \citet{samanta2013elasto} have demonstrated that the transition threshold to inertial turbulence can be controlled by varying the polymer concentration. They computed the puff lifetimes to examine the effects of viscoelasticity on transition and differentiate the EIT states from the Newtonian turbulence in a pipe flow. They found that polymer additives delay the subcritical transition to IT conforming to their drag reducing effects in turbulent channel flows. They also demonstrated that, as the shear rate is increased, a separate instability occurs at a much lower Reynolds number than the inertial turbulence and triggers a very different type of disordered motion of elasto-inertial turbulence.  \citet{foggi2024unified} showed that a reverse pathway can also be followed to achieve the EIT state. That is, viscoelasticity is introduced into an already turbulent Newtonian flow and then the Reynolds number is reduced gradually. Unlike the Newtonian flow, which would immediately re-laminarize below a certain threshold value of $Re$, the viscoelastic flow is capable of maintaining its turbulent state even at a Reynolds number as low as $Re=0.5$ in a rectilinear channel. Moreover, at a low Reynolds number, if some external perturbations are superimposed on a laminar viscoelastic flow field, these instabilities can grow and achieve the EIT state under a certain parametric setting (\citet{garg2018viscoelastic}, \citet{beneitez-etal-prf-2023,beneitez2024multistability}).

The dynamic features of EIT differ greatly from the inertial turbulence, which is the focus of many contemporary investigations. \citet{sid2018two} have demonstrated that the essential dynamic structures of EIT exist even in a two-dimensional (2D) flow. More recently, \citet{dubief2022first} have identified four distinct regimes of the stress field in a 2D channel flow by using the FENE-P viscoelastic model. These regimes were named as chaotic, chaotic arrowhead, intermittent arrowhead and stable arrowhead regimes based on constant contours of first normal stress difference ($N_1$) and pressure field. A sharp pressure gradient was observed within the flow field, similar to the one across a shock wave, promoting the shape of an arrowhead. In this 2D flow field, it was reported that an increase in the polymer concentration promoted stability while an increase in the domain length promoted chaos, indicating the role of large scales in the dynamics of EIT. The drag was found to be increasing in all these EIT regimes when compared to the corresponding laminar states. In a 3D periodic channel flow, \citet{dubief2011polymer} observed an alternating train of positive and negative $Q$-iso-surfaces (second invariant of velocity gradient tensor) near the channel wall during transition to the EIT state. Interestingly, these patterns sustained on a smaller scale unlike the inertial turbulence even when the viscoelastic flow got fully developed. These patterns were interpreted as the regions of local rotation ($Q > 0$) and dissipation of turbulent kinetic energy ($Q < 0$), and were found to change their alignment from streamwise to spanwise direction during low-drag events. Another striking difference between IT and EIT is the reversed energy cascade. Unlike the inertial turbulence where the energy transfer occurs from large to small scales, energy transfer of turbulent elastic energy to turbulent kinetic energy is observed to be from smaller to larger scales in EIT (\cite{dubief2013mechanism}).

Despite an ever increasing interest in this area in recent years, many fundamental aspects related to EIT are still elusive. In the present study, we focus on two such aspects. First, instead of providing any external perturbations to the flow, direct numerical simulations of concentrated viscoleastic flows are performed in a square-shaped channel by gradually increasing the fluid elasticity while keeping the Reynolds number fixed at a low value for which the corresponding Newtonian flow is fully laminar to examine the conditions for attaining an EIT state. Secondly, a novel mechanism of injecting  bubbles into a laminar viscoelastic flow is investigated to ascertain whether the presence of bubbles can trigger a transition to an EIT state in this straight channel. For this purpose, extensive interface-resolved direct numerical simulations are performed using a finite-difference/front-tracking method (\citet{unverdi1992front}). The log-conformation method (\citet{fattal2005time}) is employed to handle the stiff constitutive equations of the viscoelastic model (Giesekus) at high Weissenberg numbers.

The numerical method and the computational setup are described in $\S$\ref{NM} $\&$ $\S$\ref{CS}, respectively. The results are presented and discussed for a single-phase and then for the multiphase flows in $\S$\ref{Results}, followed by the conclusions in $\S$\ref{Conclusions}. 
   
\section{Governing Equations and Numerical Method}\label{NM}

\begin{figure}
\centering
  \includegraphics[width=0.9\textwidth]{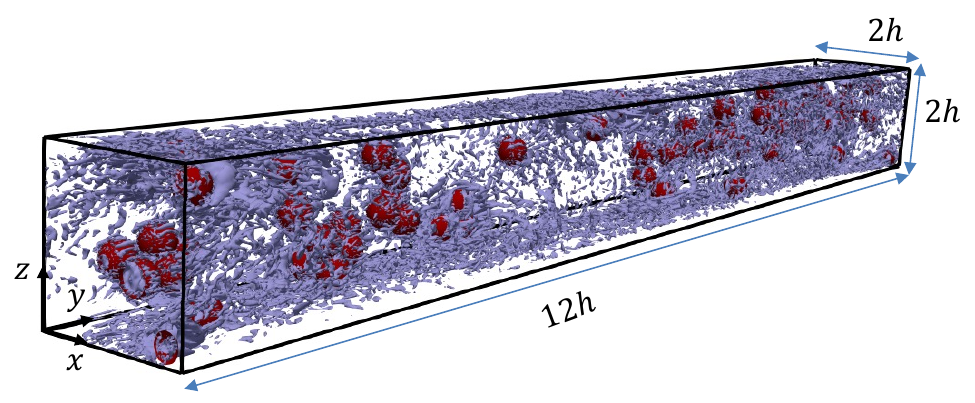}
  \caption{ The computational domain and the coordinate system considered in the present study. The constant contours of $Q$-criterion are shown around the bubbles (red colour) at a value of $Q=25$ ($Re=1000, Wi=10, Ca=0.01, \beta=0.1$).}
\label{domain}
\end{figure}

The flow equations and the Giesekus model are presented within the framework of the finite-difference/front-tracking (FT) method. The front-tracking  method used in the present study was originally developed by \cite{unverdi1992front}. \citet{izbassarov2015front} added a capability to this FT method to simulate viscoelastic two-phase systems in which one or both the phases could be viscoelastic. This robust and high fidelity method has been extensively used in our several previous works involving laminar (e.g., \citet{izbassarov2016computational}, \citet{izbassarov2016effects}, \citet{naseer2023dynamics}, \citet{Naseer_Izbassarov_Ahmed_Muradoglu_2024}) as well as  turbulent multiphase flows (e.g., \citet{ahmed2020effects}, \citet{izbassarov2021polymer}). Referring to the coordinate system shown in Fig.~\ref{domain} and employing the one-field formulation, the Navier-Stokes equations are written as

\begin{eqnarray}
\rho\frac{\partial \mathbf{u}}{\partial t}
+{\rho} \boldsymbol{\nabla}\cdot({\mathbf{u}}{\mathbf{u}})
=-\boldsymbol{\nabla}{p} -\frac{dp_o}{dy}\mathbf{j}
+\boldsymbol{\nabla}\cdot{\boldsymbol\tau} 
+\boldsymbol{\nabla}\cdot\mu_s(\boldsymbol{\nabla}{{\mathbf{u}}}+\boldsymbol{\nabla}{{\mathbf{u}}^T}) \nonumber \\ + \int_A \sigma\kappa\mathbf{n}\delta(\mathbf{x}-\mathbf{x}_f)dA,
\label{NS}
\end{eqnarray}

\noindent 
where ${\mathbf{u}}$, $\boldsymbol{\tau}$, $p$, $\rho$ and $\mu_s$ are the velocity vector, the polymer stress tensor, the pressure, and the discontinuous density and solvent viscosity fields, respectively. A pressure gradient $-\frac{dp_o}{dy}\mathbf{j}$ is applied to drive the flow and it is adjusted dynamically to keep the flow rate constant, where $\mathbf{j}$ is the unit vector in the $y$-direction. The effect of surface tension is added as a body force term on the right-hand side of the momentum equation where $\sigma$ is the surface tension coefficient, $\kappa$ is twice the mean curvature and $\mathbf{n}$ is a unit vector normal to the interface. As the surface tension acts only on the interface, $\delta$ represents a three-dimensional Dirac delta function with the arguments $\mathbf{x}$ and $\mathbf{x}_f$ being a point at which the equation is evaluated and a point at the interface, respectively. The momentum equation is supplemented by the incompressibility condition
 \begin{equation}
 \boldsymbol{\nabla} \cdot \mathbf{u} = 0.
 \label{continuity}
 \end{equation}
Viscoelasticity of bulk liquid is modelled using the Giesekus model (\cite{giesekus1982simple}). Since this model accounts for the polymer–polymer interactions, it is a suitable choice to simulate concentrated polymer solutions (\cite{varchanis2022evaluation}). In the Giesekus model, the polymer stress tensor $\boldsymbol{\tau}$ evolves by
\begin{equation}
\boldsymbol{\tau} = \frac{\mu_p}{\lambda}(\boldsymbol {B} - \boldsymbol{I}), 
\label{stress_eq}
\end{equation}
where $\mu_p$ is the polymer viscosity,  $\lambda$ is the polymer relaxation time, $\boldsymbol{B}$ is the conformation tensor and $\boldsymbol{I}$ is the identity tensor. The conformation tensor evolves by

\begin{equation}
 \frac{\partial \boldsymbol{B}}{\partial t} + \mathbf{u}\cdot\boldsymbol{\nabla}\boldsymbol{B} - \boldsymbol{\nabla}\mathbf{u}^T \cdot \boldsymbol{B} - \boldsymbol{B}\cdot\boldsymbol{\nabla}\mathbf{u} = \frac{1}{\lambda} [(1-\alpha)\boldsymbol{I} + (2\alpha - 1)\boldsymbol{B} - \alpha\boldsymbol{B^2}],
\label{conformation_tensor}
\end{equation}

\noindent where $\alpha$ is the mobility factor representing the anisotropy of the hydrodynamic drag exerted on the polymer molecules. Due to the thermodynamic considerations, $\alpha$ is restricted to $0\le \alpha\le 0.5$ (\cite{schleiniger1991remark}). When $\alpha = 0$, the Giesekus model reduces to the Oldroyd-B model. 

At high Weissenberg numbers, these highly non-linear viscoelastic constitutive equations become extremely stiff, which makes their numerical solution a challenging task. Artificial diffusion terms may be added to the discretized viscoelastic model equations to alleviate this problem. However, it has been found that, when the model equations are discretized using the conventional numerical schemes,  this artificial diffusion method may alter the mathematical nature of the constitutive equations from hyperbolic to parabolic, affecting the integrity of the simulated EIT physics, especially at high Weissenberg numbers (\cite{dubief2023elasto}). The log-conformation method offers an alternative solution to this problem and it is used in the present study. In this approach, an eigen decomposition is employed to re-write the constitutive equation of the conformation tensor (\cite{fattal2005time}, \cite{izbassarov2018computational}). It is found that, although it is computationally more expensive than the artificial diffusion method, the log-conformation retains the hyperbolic nature and thus removes the stiffness of the viscoelastic model equations, allowing robust and accurate numerical solutions by preserving large-scale features in the simulations of elastic turbulence as shown recently by \cite{Yerasi_Picardo_Gupta_Vincenzi_2024}. The interested readers are referred to \cite{fattal2005time} for the details of the procedure. 

The flow equations (Eqs.~(\ref{NS}) and (\ref{continuity})) are solved fully coupled with the Giesekus model equation (Eq.~(\ref{stress_eq})). A QUICK scheme is used to discretize the convective terms in the momentum equations while second-order central differences are used for the diffusive terms. For the convective terms in the viscoelastic equations, a $5^{th}$-order WENO-Z (\cite{borges2008wai}) scheme is used. An FFT-based solver is employed to solve the pressure Poisson equation. Since the pressure equation is not separable due to variable density in the present multiphase flow, the FFT-based solvers cannot be used directly. To overcome this challenge, a pressure-splitting technique presented by \cite{dong2012time} and \cite{dodd2014fast} is employed. A predictor-corrector scheme is used to achieve second-order time accuracy as described by \citet{tryggvason2001front}. The details of the front-tracking method can be found in the book by \cite{tryggvason2011direct} and in the review paper by \cite{tryggvason2001front}, and the treatment of the viscoelastic model equations in  \citet{izbassarov2015front} and \citet{izbassarov2018computational}. The present numerical scheme is second order accurate both in space and time. 

\section{Computational setup}\label{CS}
Figure~\ref{domain} shows the computational domain which is a square channel with the dimensions of $2h\times12h\times2h$ in the $x$, $y$ and $z$ directions, respectively, where $h$ is half of the channel width. Periodic boundary conditions are applied in the streamwise ($y$) direction whereas the other two directions ($x$ and $z$) have no-slip/no-penetration boundary conditions. The length of the channel is set to $12h$ so that a sufficient length is available for the bubbles to exhibit their effects in the viscoelastic flow field. 

The flow conditions are characterized by the following non-dimensional numbers defined as

\begin{eqnarray}
Re = \frac{\rho_o u_o h}{\mu_o},\
Wi = \frac{\lambda u_o}{h},\
Ca = \frac{\mu_o u_o}{\sigma},\
\beta = \frac{\mu_s}{\mu_o},
\label{ND_numbers}
\end{eqnarray} 

\noindent where $Re$, $Wi$ and $Ca$ are the Reynolds, Weissenberg and capillary numbers, respectively. Further, $\beta$ is the ratio of the solvent viscosity ($\mu_s$) to the zero shear viscosity ($\mu_o$) of the viscoelastic fluid. In Eq.~(\ref{ND_numbers}), $\rho_o$ is the density of the bulk fluid and $u_o$ is the average velocity in the channel. The viscosity ratio is fixed at $\beta=0.1$, representing a highly concentrated polymer solution, for all the results presented in this paper.

The present study covers the Reynolds and the Weissenberg numbers spanning in the ranges of $10 \leq Re \leq 1000$ and $1 \leq Wi \leq 1000$. Note that the highest value of $Re=1000$ is still less than the minimum value required to sustain inertial turbulence in a channel flow (\citet{owolabi2016experiments}). For these low values of $Re$, the grid resolution near the wall cannot be determined by the classical wall criterion of $x^+< 1$ and $z^+ < 1$ . As the bubbles are also injected into the flow, despite a low $Re$, a sufficiently fine grid is required to resolve the bubble interfaces and to capture the essential flow features of this chaotic viscoelastic flow regime. Therefore, a grid size of $320 \times 960 \times 320$ is used in the $x$, $y$ and $z$ directions even for $Re=10$ case. For a reference, this grid resolution is sufficient to achieve $x^+< 0.8$ and $z^+ < 0.8$ in an inertial turbulent channel flow at $Re=5600$ ($Re_\tau = 180$). For the multiphase cases, there are $64$ grid cells per equivalent bubble diameter with this selected grid size.

The flow is driven by applying $dp_o/dy$ in the negative  $y$-direction. This external pressure gradient is adjusted dynamically at each time step to keep the flow rate (and thus the bulk Reynolds number) constant in the channel. To normalize various quantities reported in this study, $h$ is used as the length scale, $u_{o}$ as the velocity scale and $h/u_{o}$ as the time scale. The normalized quantities are denoted by the superscript $^*$. The stresses are normalized by $\rho_o v_{\tau}^2$, where $v_\tau = \sqrt{\bar \tau_w / \rho_o}$ with $\bar \tau_w$ being the average total wall shear stress. Once the flow reaches a statistically steady state, the combined force due to the shear stresses at the four walls of the channel is balanced by the force due to the applied pressure gradient $dp_o/dy$.  In this multiphase flow, the density and viscosity of the bubble are denoted by $\rho_i$ and $\mu_i$, respectively. The density and the viscosity ratios are set to $\rho_{o}/\rho_{i}=10$ and $\mu_{o}/\mu_{i}=80$, respectively. These comparatively small ratios are used to enhance numerical stability and thus relaxing the time step restrictions. Apparently, this low density ratio may seem unrealistic when compared to a liquid-air system. However, a higher density ratio does not affect the bubble dynamics as has been reported by \citet{bunner2002dynamics} for a Newtonian flow. Further simulations are also performed here to check the effects of the density and viscosity ratios on the viscoelastic multiphase flows considered in this study. As documented in the Appendix, the results are not very sensitive to a further increase in the density and viscosity ratios beyond $\rho_{o}/\rho_{i}= 10$ and $\mu_{o}/\mu_{i}=80$.

\section{Results and discussion}\label{Results}

Simulations are first performed for a single-phase flow by gradually increasing the Reynolds number from $Re=10$ to $Re=1000$. Subsequently, the flow is made more viscoelastic by gradually increasing the Weissenberg number in the range of $1 \leq Wi \leq 1000$ at a fixed Reynolds number to ascertain whether the flow attains an EIT state without introducing any explicit external perturbations. To facilitate the development of second normal stress difference ($N_2$), a square-channel (duct) is used and a small shear-thinning effect is also incorporated by setting $\alpha=0.001$ for the Giesekus model. A concentrated polymer solution ($\beta=0.1$) is selected to follow the pathway identified by \citet{samanta2013elasto} for the earlier possible transition to an EIT regime. After identifying the range of $Wi$ for which the single-phase flow remains essentially laminar, bubbles are subsequently injected into fully developed viscoelastic laminar flows at $Wi=1, 5$ and $10$ while keeping $Re$ constant to ascertain whether the presence of bubbles can trigger a flow instability. The bubble volume fraction is kept constant at $3\%$ for all the cases. The capillary number is fixed at a low value of $Ca=0.01$ to maintain approximately spherical bubble shapes. The resultant single-phase and multiphase flows are presented and discussed in the following sections.
\subsection{EIT in a single-phase flow}\label{4.1}
\begin{figure}
\includegraphics[width=0.8\textwidth,right]{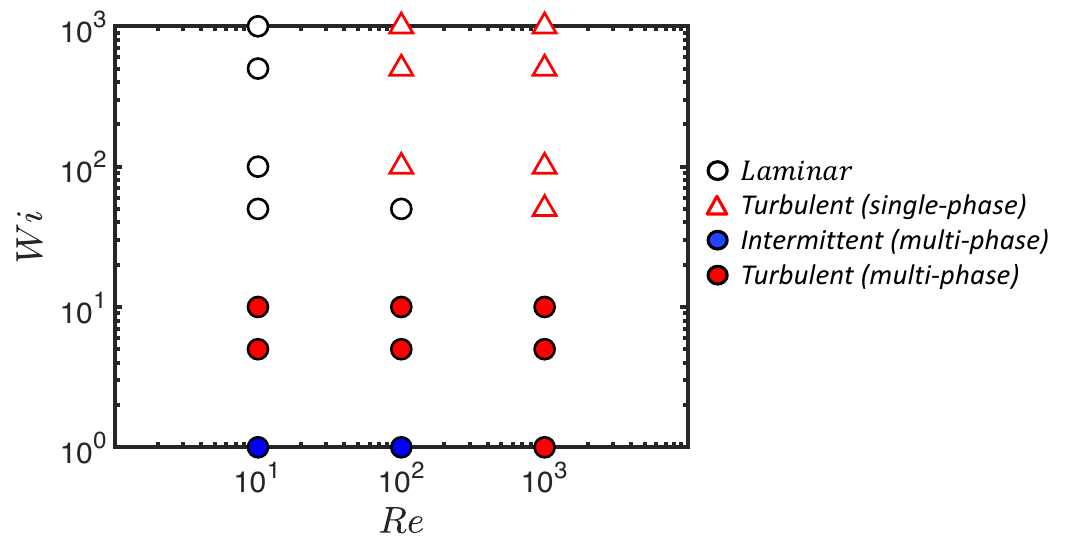}
\caption{ Various flow states in the $Re$-$Wi$ space. ($ Ca=0.01, \beta=0.1$)}
\label{phase}
\end{figure}

Figure \ref{phase} summarizes the flow states in the $Re$-$Wi$ space. At a low value of $Re=10$, the single-phase flow remains fully laminar for up to $Wi=1000$. Two points are selected in the computational domain to collect all the components of velocity and viscoelastic stresses at each time step to monitor their evolution to ascertain the attainment of a statistically steady-state solution. One of the `numerical probes' is located at the centre of the channel ($x^*$=$1$, $y^*$=$6$, $z^*$=$1$) while the other one at $25\%$ distance ($x^*$=$0.5$, $y^*$=$6$, $z^*$=$1$) from one of the walls. Once the flow reaches a statistically steady state, the simulations are continued at least for another $3\lambda$ time units, during which full flow fields are stored for a further statistical analysis.  

The simulations are then repeated for $Re=100$. While gradually increasing $Wi$, it is observed that the flow remains laminar till $Wi=100$. However, for  $Wi \geq 100$, instabilities start to appear in this low $Re$ flow. These instabilities start to grow with time, and fluctuations are observed ultimately in all three components of velocity akin to a typical turbulent signal. Viscoelastic stresses also start to fluctuate in time domain, as collected by the numerical probes at two different locations of the channel. It is important to emphasize that no explicit perturbations are provided from outside the system to initialize any instabilities in the flow. Only for $Wi\geq100$, the viscoelastic stresses  are high enough to trigger an instability that ultimately leads to a chaotic flow regime. As the flow rate is kept constant for a fixed $Re$, a sudden increase in the applied pressure gradient is also observed once the flow is transitioned from laminar to this chaotic state, indicating a rapid increase in the drag. The characteristics of this chaotic flow regime for $Wi \geq 100$ will be discussed in detail in the subsequent paragraphs.

When $Re$ is increased further to $1000$, the transition to a chaotic flow regime starts to appear at a lower value of $Wi \geq 50$. This $Re$ is still smaller than the minimum $Re_{cr} \approx 1100$ required for sustaining the inertial turbulence in a channel flow (\citet{owolabi2016experiments}). As $Wi$ is increased further, the turbulent intensity of this chaotic regime increases with a change in statistical quantities as well.

\begin{figure}
\centering
\includegraphics[width=1.0\textwidth]{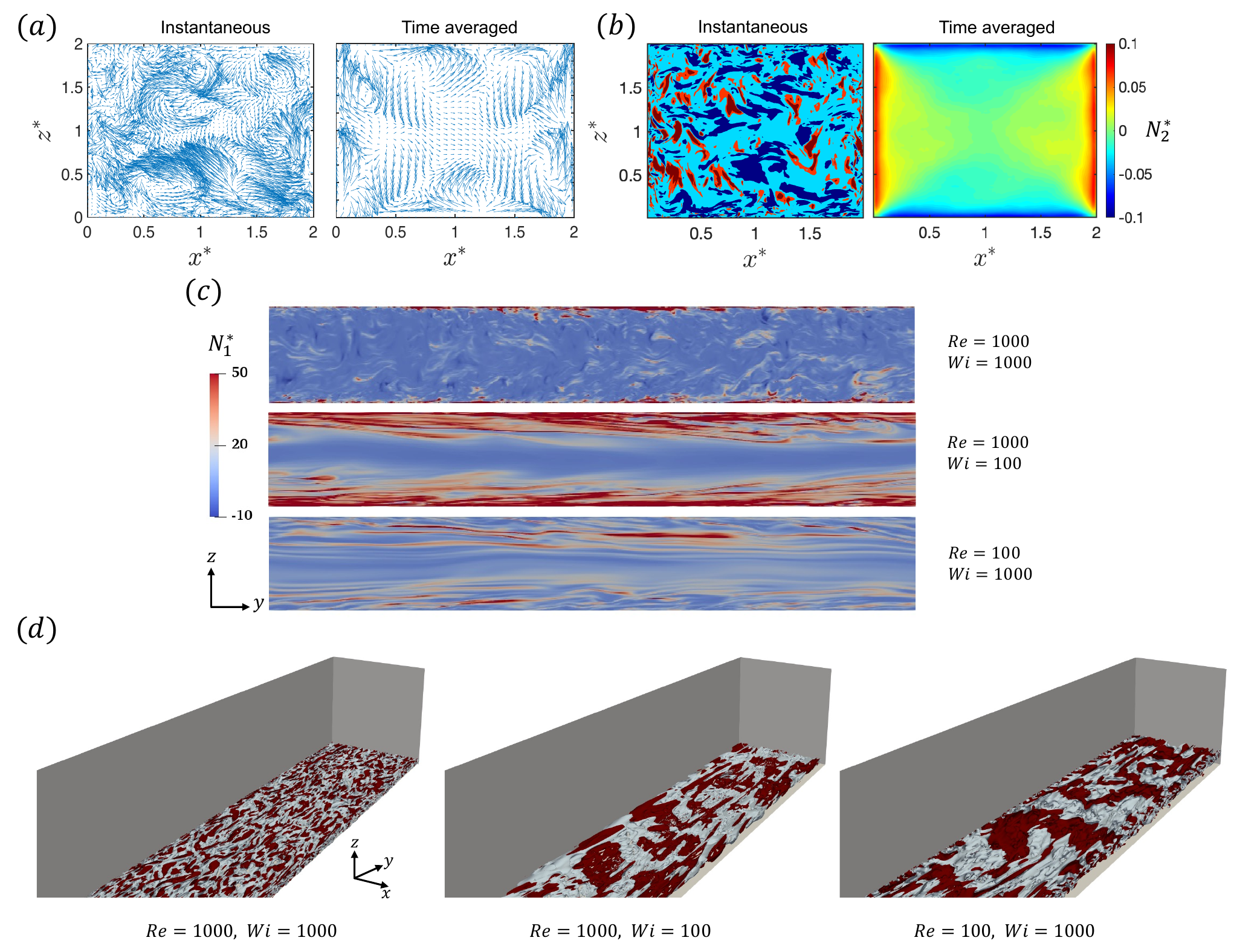}
\caption{\textit{(a)} Instantaneous (\textit{left}) and time-averaged (\textit{right}) snapshots of velocity vectors showing the secondary flow field at $Re=1000, Wi=1000$ \textit{(b)} Instantaneous (\textit{left}) and time-averaged (\textit{right}) snapshots of second normal stress difference ($N_2$) at $Re=1000$ and $Wi=1000$ \textit{(c)}. Contours of first normal stress difference ($N_1$) are shown in a statistically steady EIT regime for different values of $Re$ and $Wi$. \textit{(d)} Iso-surfaces of $Q$-criterion are shown close to the bottom wall of the channel in a statistically steady EIT regime for different values of $Re$ and $Wi$. The red and gray colours represent the positive and the negative values, respectively. The contours are plotted at $\pm{5}$, $\pm{0.05}$ and $\pm{0.005}$ for ($Re, Wi$) = ($1000, 1000$), ($1000, 100$) and ($100, 1000$) cases, respectively.}
\label{sp-contours}
\end{figure}

Figure~\ref{sp-contours}a shows the velocity vectors in a vertical cutting $xz$-plane of the viscoelastic duct flow once it reaches a statistically steady state for the $Re=1000$ and $Wi=1000$ case. The instantaneous and time-averaged flow fields are plotted in left and right panels of Fig.~\ref{sp-contours}a, respectively. As seen, the instantaneous flow field exhibits a chaotic flow regime while the time-averaged velocity field shows an almost similar secondary flow pattern with eight vortices as also observed in the viscoelastic laminar duct flow with the shear-thinning effect (\citet{li2015dynamics}, \citet{Naseer_Izbassarov_Ahmed_Muradoglu_2024}). This secondary flow field is developed in the duct flow primarily due to the presence of a second normal stress difference $N_2 = \bar\tau_{zz} - \bar\tau_{xx}$. The instantaneous and time-averaged contour plots of $N_2$ are shown in the left and right panels of Fig.~\ref{sp-contours}b, respectively. The instantaneous plot confirms the chaotic flow pattern while the time-averaged one resembles the typical distribution of $N_2$ in a laminar duct flow. As the EIT regime is associated with long-elongated sheets of the first normal difference ($N_1$), the distribution of $N_1$ is also shown in Fig.~\ref{sp-contours}c in the $yz$-plane at the centre of the duct for the various combinations of $Re$ and $Wi$ for which the single-phase flow becomes chaotic. \citet{dubief2023elasto} categorized different EIT regimes based on the structures of $N_1$ for a 2D channel flow. For a low value of $Re=100$, a similar `chaotic regime' is observed in the present 3D duct flow as shown in Fig. \ref{sp-contours}c. At a higher value of $Re=1000$ but with a moderate value of $Wi=100$, this regime is shifted to `chaotic arrowhead' and once the Weissenberg number is increased to a very high value of $Wi=1000$, the long-elongated sheets of $N_1$ break down completely and the flow starts to resemble a typical inertial turbulence. The iso-surfaces of the second invariant of velocity gradient tensor ($Q$-criterion) is depicted in Fig.~\ref{sp-contours}d to show the corresponding change in the near-wall vortices for the same combinations of $Re$ and $Wi$. At a very high value of $Re$ and $Wi$, the positive and negative contours of the $Q$-criterion show a chaotic pattern near the wall. With decreasing $Re$ or $Wi$, the spanwise elongated pattern of positive and negative contours of the $Q$-criterion starts to appear indicating that the signatures of elastic turbulence become more and more apparent in the EIT regime once the elastic effects dominate over the inertial effects. In the absence of any perturbations from outside the system, the flow is laminar at a lower value of $Wi$ in the present scenario, as shown in Fig.~\ref{phase}. Therefore, a clear pattern of alternating contours of $Q$-criterion is not visible here as also observed by \citet{dubief2023elasto}.

\begin{figure}
\centering
\includegraphics[width=1.0\textwidth]{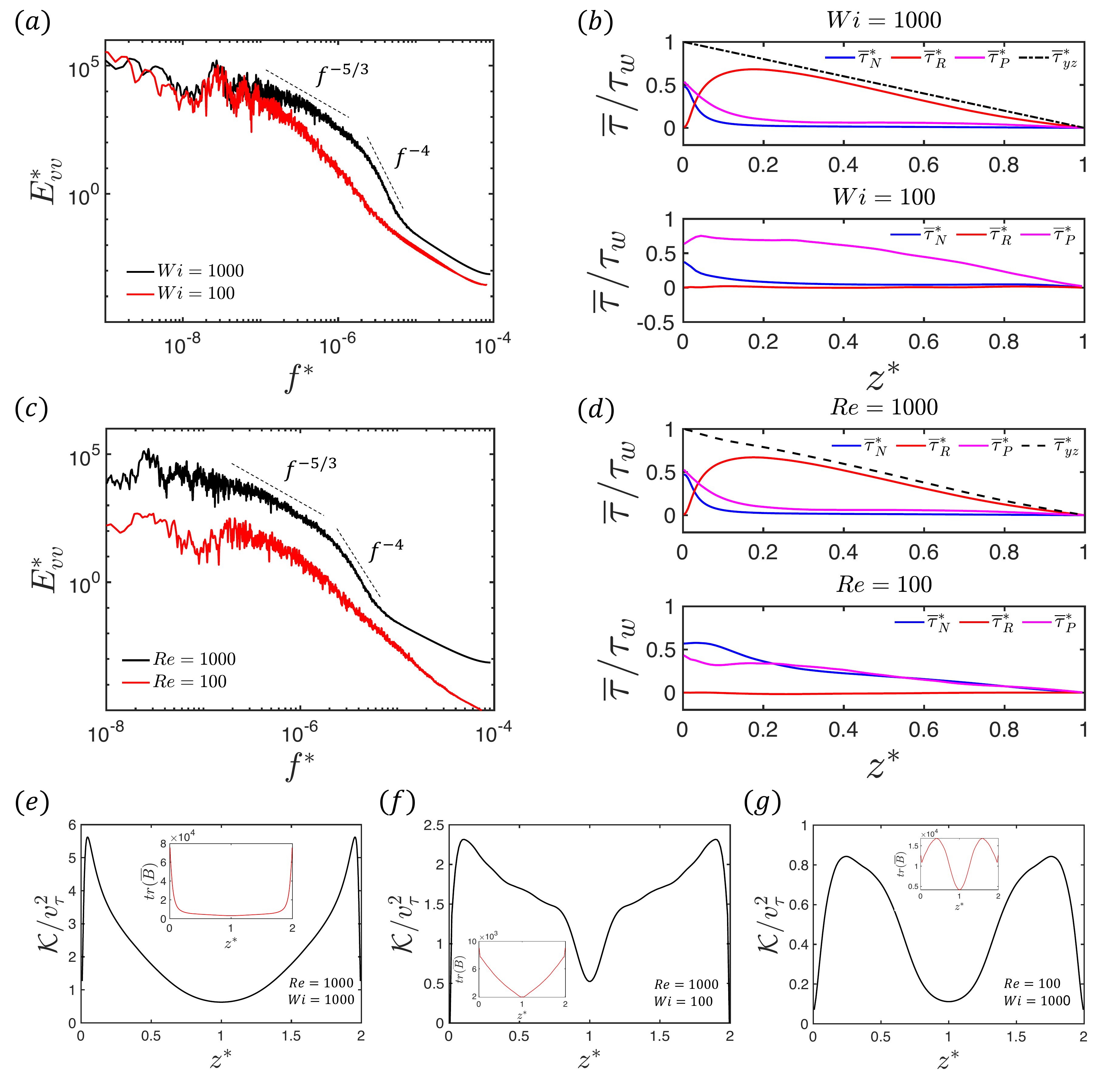}
\caption{The 1D energy spectrum for \textit{(a)} $(Re,Wi)=(1000,1000)$ \& $(Re,Wi)=(1000,100)$ and for \textit{(c)} $(Re,Wi)=(1000,1000)$ \& $(Re,Wi)=(100,1000)$. The distribution of the components of the average shear stress from the channel wall towards the centre in the mid-plane for \textit{(b)} $(Re,Wi)=(1000,1000)$ (top panel) \& $(Re,Wi)=(1000,100)$ (bottom panel) and for \textit{(d)} $(Re,Wi)=(1000,1000)$ (top panel) \& $(Re,Wi)=(100,1000)$ (bottom panel) \textit{(e, f, g)} The distribution of turbulent kinetic energy  in the mid-plane of the channel for different values of $Re$ and $Wi$. In the insets of respective figures, the average elongation of polymer molecules represented by $tr(\bar B)$ are shown for the same mid-planes.}
\label{single-phase}
\end{figure}

Figure~\ref{single-phase} shows various quantities used to evaluate turbulent characteristics of this single-phase chaotic flow regime for various combinations of $Re$ and $Wi$. The first one is the 1D energy spectrum plotted for $Wi=100$ and $Wi=1000$ at $Re=1000$ (Fig.~\ref{single-phase}a). The energy axis is normalized by $\frac{1}{2}\rho_o u_o^2$ and the frequency axis by $u_o/h$. At the higher value of $Wi=1000$, the kinetic energy spectrum shows a typical $-5/3$ scaling at the large scales but it shifts to $-4$ at the smaller scales (dissipation range). These values are consistent with the results obtained in a single-phase channel flow at lower values of $Re$ by \citet{mukherjee2023intermittency} and \citet{foggi2024unified}. Once $Wi$ is decreased to $Wi=100$ while keeping $Re$ the same, the scaling of energy spectrum starts to deviate from $-4$. At this comparatively higher value of $Re$ (but still lower than the minimum one required for the inertial turbulence to sustain in a Newtonian flow), the turbulent characteristics exhibit the manifestation of both the inertial and the elastic effects. Figure~\ref{single-phase}c shows the energy spectra obtained for $Re=100$ and  $Re=1000$ at $Wi=1000$. As seen, the flow is dominated by the elastic effects at $Re=100$. Hence, the maximum range of energy spectrum gets closer to the scaling of $-4$. It can be inferred from these scales that once the turbulent flow is entirely governed by inertia (classical inertial turbulence), the energy spectrum assumes a slope of $-5/3$ whereas once the turbulent regime is dominated by the elastic effects with negligible inertia, as in ET, the spectrum exhibits a slope of $-4$. Any combination of $Re$ and $Wi$ in between these two regimes assumes a slope in between these two limits.

The total mean shear stress in the mid-plane comprises three components: the viscous shear due to mean flow $\bar \tau_{\scaleto{N}{3.5pt}}$, the Reynolds stress due to velocity fluctuations $\bar \tau_{\scaleto{R}{3.5pt}}$ and a polymeric stress $\bar \tau_{\scaleto{P}{3.5pt}}$. Once the flow reaches a statistically steady state, the applied pressure gradient must balance the total shear stress at the four walls of the channel. The three components of shear stress ($\bar \tau_{\scaleto{N}{3.5pt}} = \mu_o \frac{\partial \bar v}{\partial z}$, $\bar \tau_{\scaleto{R}{3.5pt}} = -\rho_o \overline{v'w'}$, $\bar \tau_{\scaleto{P}{3.5pt}} = \bar \tau_{yz}$) are plotted for the central plane of the duct ($x^*=1$) for different values of $Re$ and $Wi$ (Figs.~\ref{single-phase}$b$ \& $d$). The fluctuating and the time-averaged quantities are denoted by prime $(.')$ and overbar $(\bar{.})$, respectively. Note that the averaging is performed both in time and in streamwise direction since the flow is expected to be homogeneous in the streamwise direction once a statistically steady state is reached. The stress components are normalized by the local shear stress at the wall of the respective plane. For high values of $Re=1000$ and $Wi=1000$, it is observed that the stress balance in the mid-plane of the channel resembles that of a classical inertial turbulence. The viscous stress is maximum at the wall and decays to zero at the channel centre whereas the Reynolds stress is zero at the wall, gradually increases to its peak value, and then goes to zero at the channel centre. The viscoelastic stress also follows the viscous stress profile and becomes maximum at the wall and zero at the channel centre. As a result, the total shear stress decays linearly from the wall towards the channel centre (Fig.~\ref{single-phase}$b$). Once $Wi$ is reduced while keeping $Re=1000$, the magnitude of Reynolds stress becomes negligibly small compared to viscous and polymeric stresses. Although the same scaling of the energy spectrum and the turbulent signal are indicative of a turbulent flow field at this comparatively lower value of $Wi=100$, the Reynolds stress becomes negligibly small compared to the overall shear stress in the mid-plane. These not-so-smooth profiles of different shear stress components indicate a clear departure of the turbulent flow field from the inertial one. A similar situation is observed at lower inertia as well. For instance, at a low Reynolds number ($Re=100$) and a high Weissenberg number ($Wi=1000$), the turbulent flow field shows almost the same scaling of energy spectrum but the total shear stress remains to be dominated by the viscous and polymer stress components (Figs.~\ref{single-phase}$d$). The distribution of the turbulent kinetic energy (TKE) $\cal K$ $= 0.5(\overline{u'^2} + \overline{v'^2} + \overline{w'^2})$ is shown in Figs. $\ref{single-phase}e$, $f$ \& $g$ at the mid-plane where TKE is found to be maximum near the walls and decays to a minimum value at the centre. Interestingly, the average elongation of polymer molecules quantified by the trace of the conformation tensor $tr(\bar{B})$ shows a similar trend for different values of $Wi$ as shown in the insets of Figs.~\ref{single-phase}$e$, $f$ \& $g$, confirming the fact that turbulent kinetic energy is mainly produced by the elastic effects in this EIT regime. It is worth mentioning that the value of the applied pressure gradient decreases as $Wi$ increases for a constant value of $Re$, indicating a decrease in the drag force. This trend is consistent with the reversed pathway; that is, when polymer molecules are added into an already turbulent Newtonian flow, drag is reduced with an increase in $Wi$ (\citet{serafini2022drag}).

It is important to highlight that exceptionally high Weissenberg numbers required to trigger flow instability in this rectilinear channel may seem impractical. However, it should be noted that $Wi$ is defined based on the average flow velocity in the channel. As shear rate decreases significantly towards the channel centre in this duct flow, the effective local Weissenberg number is actually much lower than the nominal value of $Wi$, which is based on the average flow velocity.
\subsubsection{Turbulent kinetic energy}
The turbulent kinetic energy (${\cal K}=\frac{1}{2} \overline{u'_i u'_i}$) in a viscoelastic flow evolves by

\begin{eqnarray}
 \frac{\partial {\cal K}}{\partial t} &&= 
 \overbrace{-\bar u_j \frac{\partial {\cal K}}{\partial x_j}}^{\cal A} - 
 \overbrace{\frac{1}{2}\frac{\partial (\overline{u_i' u_i' u'_j})}{\partial x_j}}^{\cal Q} - 
 \overbrace{\frac{1}{\rho_o} \overline {u'_i \frac{\partial p'}{\partial x_i}}}^{\cal R} - 
 \overbrace{\left(\overline {u'_i u'_j} \frac{\partial \bar u_i}{\partial x_j}\right)}^{\cal P} + 
 \overbrace{\frac{1}{2}\frac{\partial}{\partial x_j}(\nu_s \frac{\partial u'_i u_i'}{\partial x_j})}^{\cal D}-
\overbrace{\nu_{s} \overline{\frac{\partial u'_i}{\partial x_j}\frac{\partial u'_i}{\partial x_j}}}^{\mathlarger{\cal \epsilon}}
 \nonumber \\
 \nonumber \\
  &&+\underbrace{\frac{1}{\rho_o}\left(\overline{u'_i\frac{\partial \tau'_{ij}}{\partial x_j}}\right)}_{{\cal W}_p}
  + \underbrace{\overline{u'_i f'_i}}_{\cal B} ,\
\label{TKE}
\end{eqnarray}

\noindent where ${\cal A}$,  ${\cal Q}$, ${\cal R}$, $\cal{P}$, $\cal{D}$,  $\mathlarger{\cal \epsilon}$, ${\cal W}_{p}$ and $\cal{B}$ represent advection by mean flow,  transport by  velocity fluctuations, transport by pressure, production by  mean flow,  viscous diffusion,  viscous dissipation,  polymer work and work done by  body force, respectively. In a viscoelastic turbulent flow, ${\cal W}_{p}$ may change its sign and can serve as either dissipation or production depending upon the signs of the polymer stress fluctuations and fluctuating velocity gradients as has been observed in both experimental (\citet{ptasinski2001experiments}) and numerical (\citet{dallas2010strong}) works.

\begin{figure}
\centering
\includegraphics[width=1.0\textwidth]{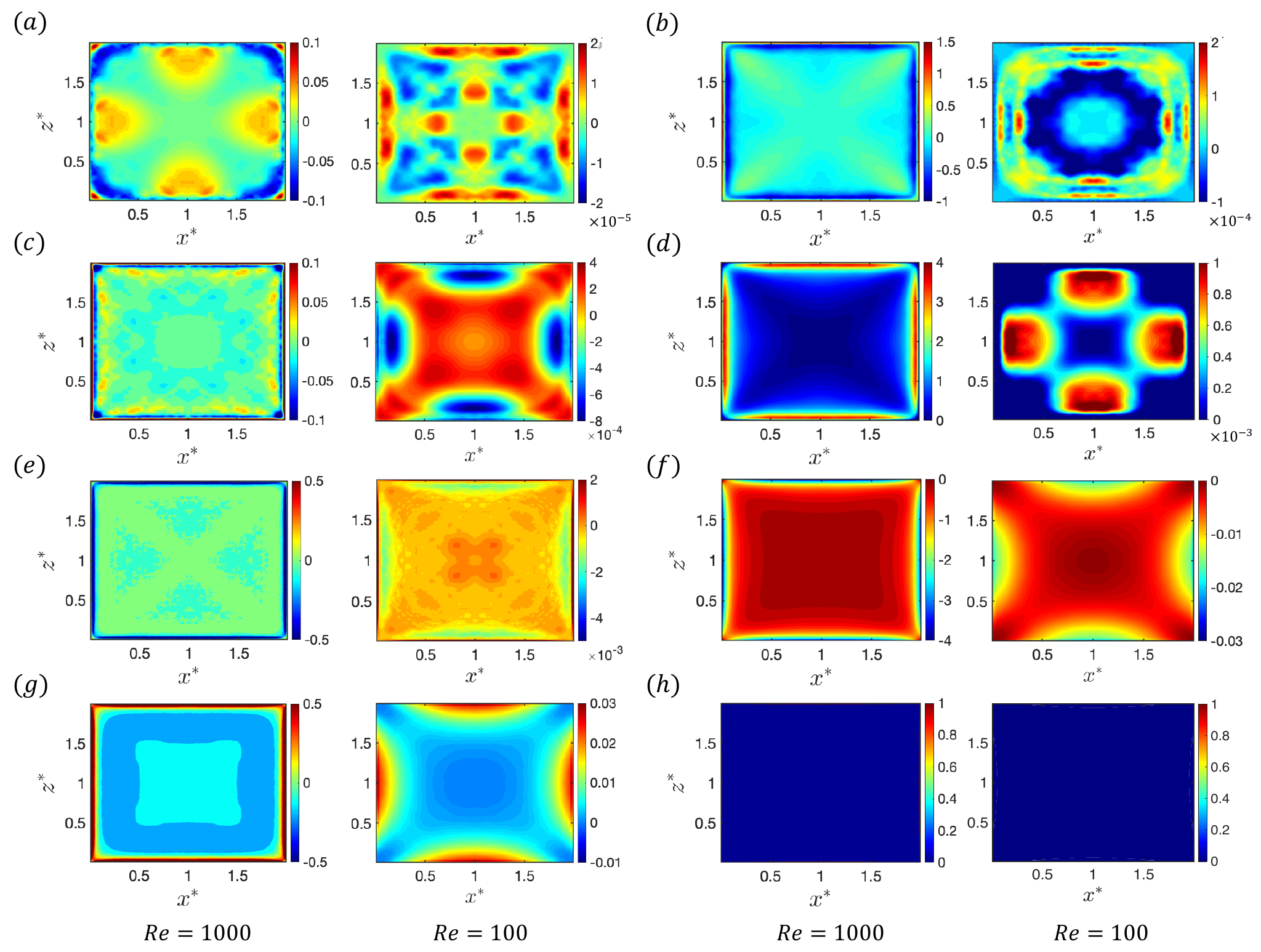}
  \caption{Contours of (\textit{a}) advection by mean flow, (\textit{b}) transport by velocity fluctuations, (\textit{c}) transport by pressure, (\textit{d}) production by mean flow, (\textit{e}) viscous diffusion, (\textit{f}) viscous dissipation, (\textit{g}) polymer work and (\textit{h}) time derivative of TKE are shown in a vertical cutting $xz$-plane for (\textit{left}) $Re=1000, Wi=1000$ and (\textit{right}) $Re=100, Wi=1000$ cases, respectively.}
\label{TKE_sp}
\end{figure}

To get better insight of the EIT regime in this duct flow, these terms are averaged in time and in the streamwise ($y$) direction once the flow reaches a statistically steady state. The constant contours of all the terms in Eq.~(\ref{TKE}) are shown in Fig.~\ref{TKE_sp} at two different values of $Re$ in a vertical cutting $xz$-plane except for the body force term that is zero for this single-phase flow. For $Re=1000$, the advection by mean flow (Fig.~\ref{TKE_sp}a-left) shows small regions of positive peaks at the four corners and towards the centre of the duct. A pattern of four negative peaks is also observed slightly away from the walls at this high Reynolds number EIT flow. However, once the Reynolds number is lowered to  $Re=100$, the magnitude of the advection term becomes negligibly small. Similarly, a symmetric pattern of transport term is observed for the higher Reynolds number (Fig.~\ref{TKE_sp}b-left) and the same term becomes negligibly small at lower inertia. At $Re=1000$, the positive and negative peaks of the transport term remain closer to the channel walls except at the corners where it remains approximately zero. An opposite trend is observed for the pressure term (Fig.~\ref{TKE_sp}c). Once the flow is dominated by the inertial effects ($Re=1000$), the pressure term remains significant while, in an elastically dominated EIT regime ($Re=100$), it shows a symmetric pattern of positive and negative zones, however, with a negligibly small magnitude. The positive peaks of the pressure term extend from four corners of the duct till the centre with negative peaks in the vicinity of the channel walls. Turbulence production by the mean flow is mainly produced near the walls except at the four corners (Fig.~\ref{TKE_sp}d) and becomes zero in the most of the central region of the duct with a symmetric pattern at $Re=1000$. However, in the elastically dominated EIT flow ($Re=100$), this pattern changes with four regions of positive peaks away from the walls. As the turbulence is mainly produced by the elastic effects at $Re=100$, the magnitude of the production term also remains negligibly small. Diffusion term (Fig.~\ref{TKE_sp}e) shows positive peaks adjacent to the walls except at the corners, immediately followed by the negative peaks before decaying to zero for $Re=1000$ whereas, for the $Re=100$ case, it is positive near the walls and remains chaotic in the most of the remaining central portions of the duct. The dissipation term shows a similar pattern for both the cases (Fig.~\ref{TKE_sp}f), i.e., it is maximum at the walls except in the corners and becomes zero towards the duct centre. The most important term in this EIT regime is the work done by the elastic force (Fig.~\ref{TKE_sp}g) as it can contribute both towards generation or dissipation of turbulence. At higher inertia ($Re=1000$), the positive peak of the polymer work is observed to be closer to the four walls of the channel. It indicates that the viscoelastic stresses produce turbulence away from the walls but act as a dissipating force closer to the walls of the channel. The same term remains negligible at the centre. Interestingly, at lower inertia where the elastic effects dominate the EIT regime, the polymer work contributes towards turbulence production at the four corners and towards the central region of the duct. The dissipative role of polymer work is mainly observed at the four walls except the corners. The viscous dissipation and the pressure terms follow the same pattern, as seen in Figs.~\ref{TKE_sp}f \& c, respectively. Finally, the summation of all the terms in Eq.~(\ref{4.1}) approaches zero confirming that a statistically steady state is reached, as shown in Fig.~\ref{TKE_sp}h.

Notably, the distribution of various components of TKE in the present EIT regime of a highly concentrated polymer solution at low $Re$ is markedly different than that in a dilute viscoelastic turbulent duct flow at high $Re$ as studied by \citet{shahmardi2019turbulent}, where the turbulent statistics of the flow were entirely dominated by the inertial effects. 

\subsection{EIT in a multiphase flow}\label{4.2}

\begin{figure}
\centering
\includegraphics[width=1.0\textwidth]{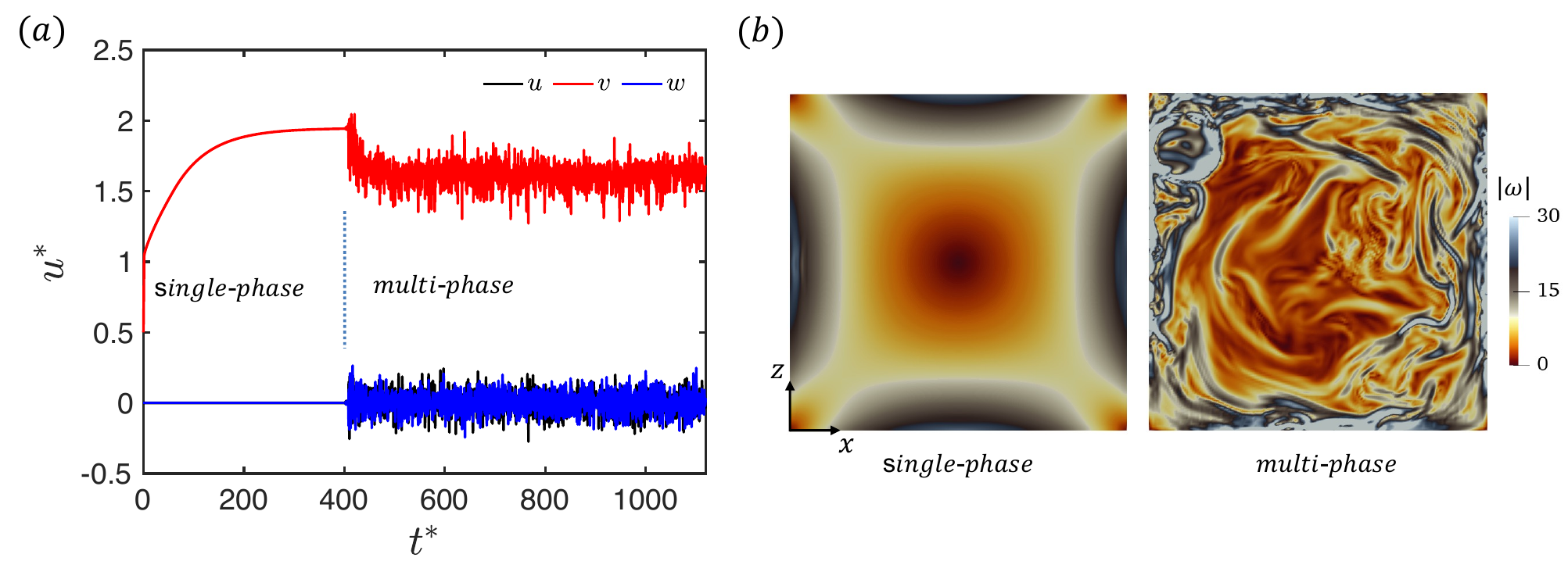}
\caption{(\textit{a}) Evolution of flow velocity and its transition to turbulent state once bubbles are injected into the flow (\textit{b}) The contours of vorticity magnitude in an $xz$-plane at the centre of the domain for the single-phase and multi-phase regimes ($Re=1000, Wi=5, \beta=0.1, Ca=0.01$)}
\label{transition}
\end{figure}

For a given Reynolds number, a significantly high $Wi$ is always required to achieve the EIT state in a single phase flow in the absence of any external perturbations, as has been shown in the previous section. When $Re$ is low (e.g., $Re=10$), even the Weissenberg number as high as $1000$ cannot make the flow unstable (Fig.~\ref{phase}). Similarly, the single-phase flow is found to be essentially laminar at the lower values of $Wi$ even when Reynolds number is as high as $Re=1000$. 

We next examine the effects of bubble injection into the viscoelastic laminar channel flow. For this purpose, simulations are performed for three relatively low values of $Wi=1$, $5$ and $10$ at each value of $Re=10$, $100$ and $1000$. Note that the the single phase flow remains laminar for these combinations of $Wi$ and $Re$. Calculations are first carried out for the single-phase flow until a statistically steady state is reached. Then, spherical bubbles are injected instantaneously into the flow with a volume fraction of $3\%$. The bubbles are initially injected randomly and uniformly in the entire duct. The capillary number is fixed at $Ca=0.01$ to keep the bubbles nearly spherical. The state of this multiphase flow field is continuously monitored at each time step using the two numerical probes as for the single-phase case. By introducing the bubbles into the flow, the flow is found to achieve a chaotic state even for the Reynolds and Weissenberg numbers as low as $Re=10$ and $Wi=5$. Time histories of instantaneous velocity components are plotted in Fig.~\ref{transition}a for the $Re=1000$ and $Wi=5$ case to show the transition from a laminar to an EIT state after the injection of bubbles. Bubbles are injected at about $t^*=400$. As seen, the signals are very smooth until the injection of bubbles indicating a laminar flow. Once the bubbles are injected into the flow, all three components of the velocity exhibit random fluctuations indicating a transition to the EIT state. These fluctuations grow with the passage of time and ultimately the entire flow becomes fully turbulent. This turbulent flow state is visualized by plotting constant contours of  vorticity ($\boldsymbol{\omega}=\nabla\times\boldsymbol{u}$) magnitude in Fig.~\ref{transition}b for the single-phase and multiphase regimes in a vertical cutting $xz$-plane at the centre of the duct. Once the flow reaches a statistically steady state after the injection of bubbles, the simulations are continued at least for another $10$ flow-through time units ($10\times 12h/u_o$) to collect data for further statistical analysis. 
\begin{figure}
\centering
\includegraphics[width=1.0\textwidth]{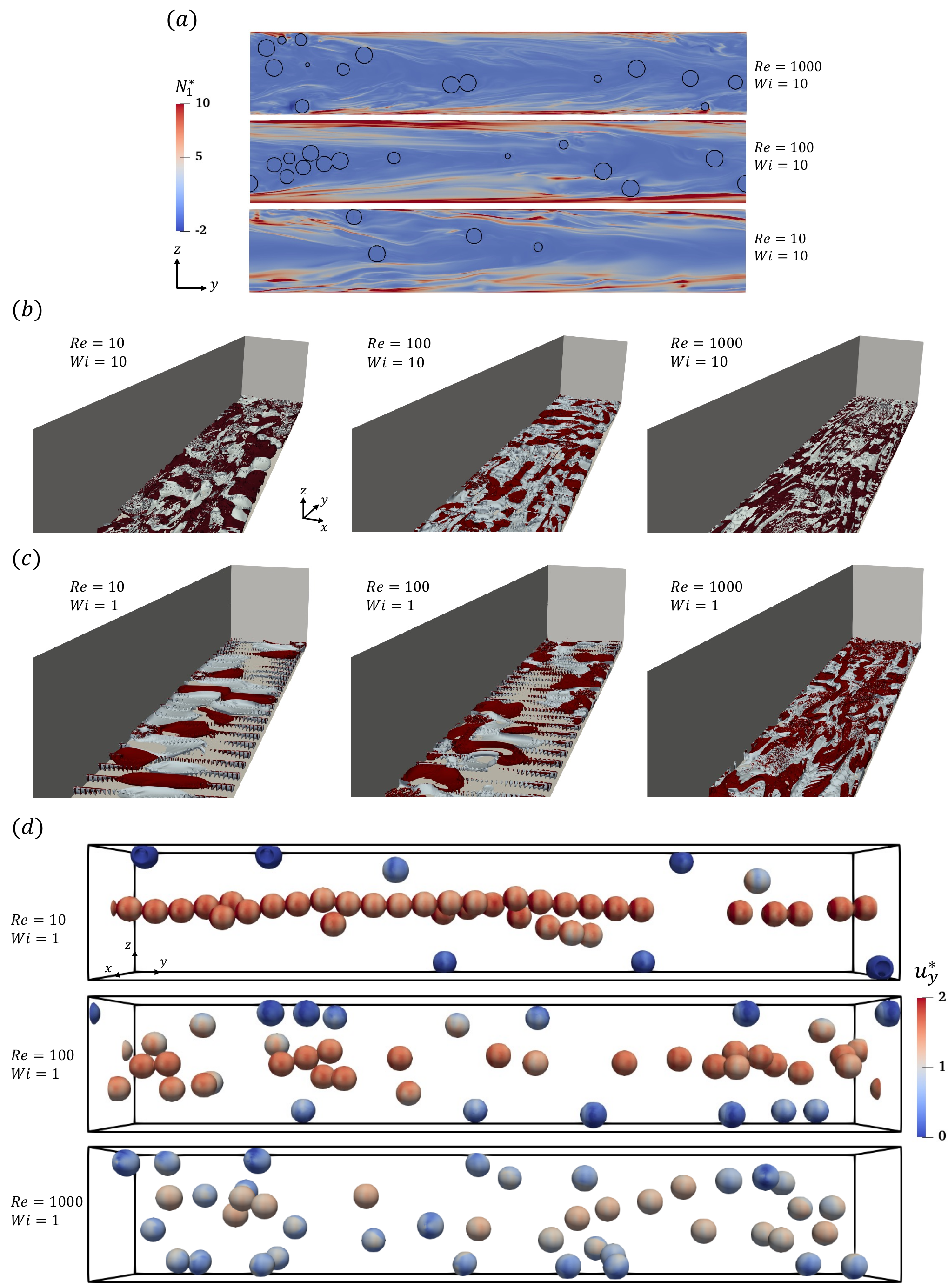}
\caption{(\textit{a}) Contours of $N_1$ in the mid-plane for different values of $Re$ at $Wi=10$. Iso-surfaces of $Q$-$criterion$ at the bottom walls of the channel for the multiphase EIT regime at (\textit{b}) $Wi=10$ and at (\textit{c}) $Wi=1$ for different values of $Re$. For (\textit{b}), the contours are plotted at $\pm{0.00001}$, $\pm{0.001}$ and $\pm{0.01}$ for $Re=10, 100$ $\&$ $1000$ cases with red colour (positive) and grey (negative), respectively. For (\textit{c}), the contours are plotted at $\pm{0.00001}$ for $Re=10, 100$ and at $\pm{0.1}$ for $Re=1000$ case. (\textit{d}) Distribution of bubbles coloured by the magnitude of streamwise flow velocity in the channel for different values of $Re$ at $Wi=1$}
\label{mp-contours}
\end{figure}

Figure~\ref{mp-contours}a shows the distribution of first normal stress difference ($N_1$) in the mid-plane for three different values of $Re=10$, $Re=100$ and $Re=1000$ at $Wi=10$. A `chaotic arrowhead' regime is observed for the all three values of $Re$ at this relatively low $Wi$ for which the single-phase flow remains fully laminar. Note that all bubbles are of the same size in Fig.~\ref{mp-contours}a and the seemingly different appearance of the bubble sizes is simply caused by the out-of-plane bubbles at this particular time instant. The constant contours of $Q$-criterion at the walls for the same three cases are depicted in Fig.~\ref{mp-contours}b, showing a chaotic pattern similar to the one observed in the single-phase cases at low value of $Wi$ (Fig.~\ref{sp-contours}d). As the instability is solely triggered by the presence of bubbles, the spanwise elongated structures of $Q$-criterion observed by \citet{dubief2023elasto} by superimposing the isotropic turbulence as an initial condition to achieve the EIT regime are not observed in the present scenario. The contours of $Q$-criterion at the wall are shown in Fig.~\ref{mp-contours}c for $Re=10$, $100$ and $1000$ at $Wi=1$. It is observed that, for this low $Wi=1$, the flow remains laminar at the walls for $Re=10$ and $Re=100$ cases. The chaotic flow regime remains confined to the central region of the duct away from the walls. The positive and negative structures of $Q$-criterion are only visible at the walls once plotted at negligibly smaller values (i.e; $\pm 1\times10^{-5}$). For $Re=1000$, however, a chaotic flow regime is observed even at $Wi=1$ as seen by the contours of $Q$-criterion (Fig.~\ref{mp-contours}c).  

\begin{figure}
\centering
\includegraphics[width=0.9\textwidth]{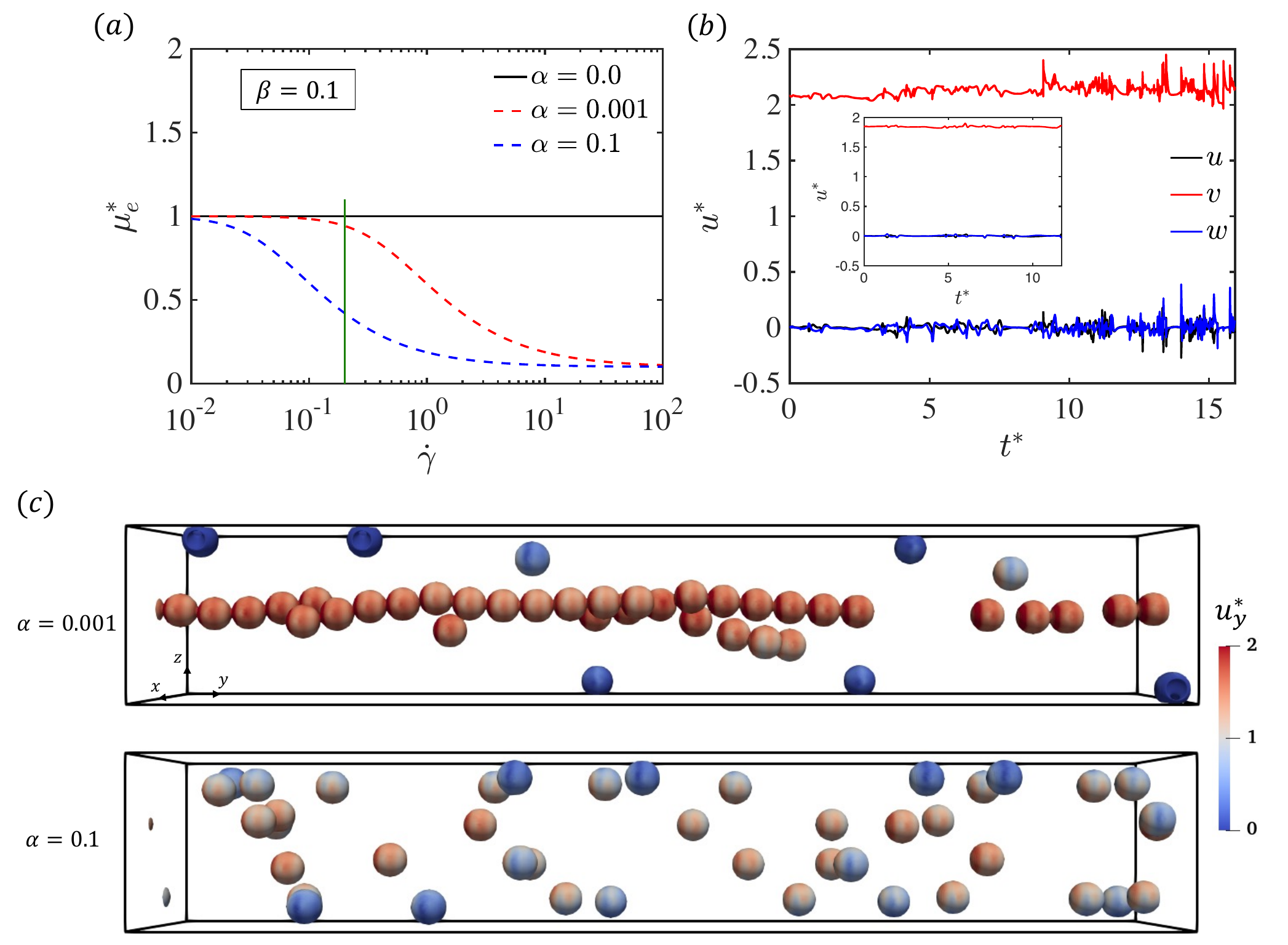}
  \caption{(\textit{a}) The change in the effective viscosity of a Giesekus fluid in a concentrated polymer solution. The vertical line in (\textit{a}) shows the value of shear rate at the wall for the central yz-plane of the duct. (\textit{b}) The velocity signal collected from the central point ($x^*=1, y^*=6, z^*=1$) of the domain for $\alpha=0.001$. In the inset, the same signal is shown for $\alpha=0.1$ case. (\textit{c}) The distribution of bubbles in the duct for different values of shear thinning parameter $\alpha$ ($Re=10, Wi=1, Ca=0.01, \beta=0.1$)}
\label{alignment}
\end{figure}

The bubble distributions in the duct are depicted in Fig.~\ref{mp-contours}d. It is interesting to see that, although the bubble distribution is essentially uniform in the other cases, the majority of the bubbles are aligned in the centre of the duct forming a string-shaped pattern for the lower values of $Re=10$ and $Wi=1$, which also explains the reason that the disturbances remain confined to the core region of the duct away from the walls for this case. In the case of non-colloidal spherical solid particles, \citet{won2004alignment} have experimentally shown that although the main driving force for the lateral migration of particles is the first normal stress difference, particle alignment is actually promoted by the shear-thinning effect for a particular rheological setting. In the present case of bubbles, the role of shear-thinning effect in the alignment of bubbles is investigated by performing another simulation at $Re=10$ and $Wi=1$ with a higher shear thinning and the results are shown in Fig.~\ref{alignment}. The change in the effective viscosity ($\mu_e$) of the flow due to shear-thinning effect at different shear rates ($\Dot{\gamma}$) is governed by the concentration of polymers ($\beta$) and the mobility factor ($\alpha$) in the Giesekus model (Fig.~\ref{alignment}a). The slope of viscosity versus shear rate plot is controlled by the mobility factor $\alpha$ and the final value of viscosity at a high enough shear rate is determined by the concentration of polymers $\beta$ (\citet{yoo1989steady}). The vertical line in Fig.~\ref{alignment}a indicates the shear rate at the wall for the mid-plane of the duct for the $Re=10$ and $Wi=1$ case. Once the mobility factor is increased to $\alpha=0.1$, a random distribution of bubbles is observed instead of the string-shaped aggregation at the centre of the duct (Fig.~\ref{alignment}c). We thus conclude that, unlike the solid particles, a higher shear-thinning effect of the ambient fluid breaks up the alignment of bubbles at least for this particular rheological setting. Further investigation is required to reveal the exact mechanism behind the role of shear-thinning effect on different forces acting on the bubbles and causing them to align in the central region of the duct. 

\begin{figure}
\centering
\includegraphics[width=0.97\textwidth]{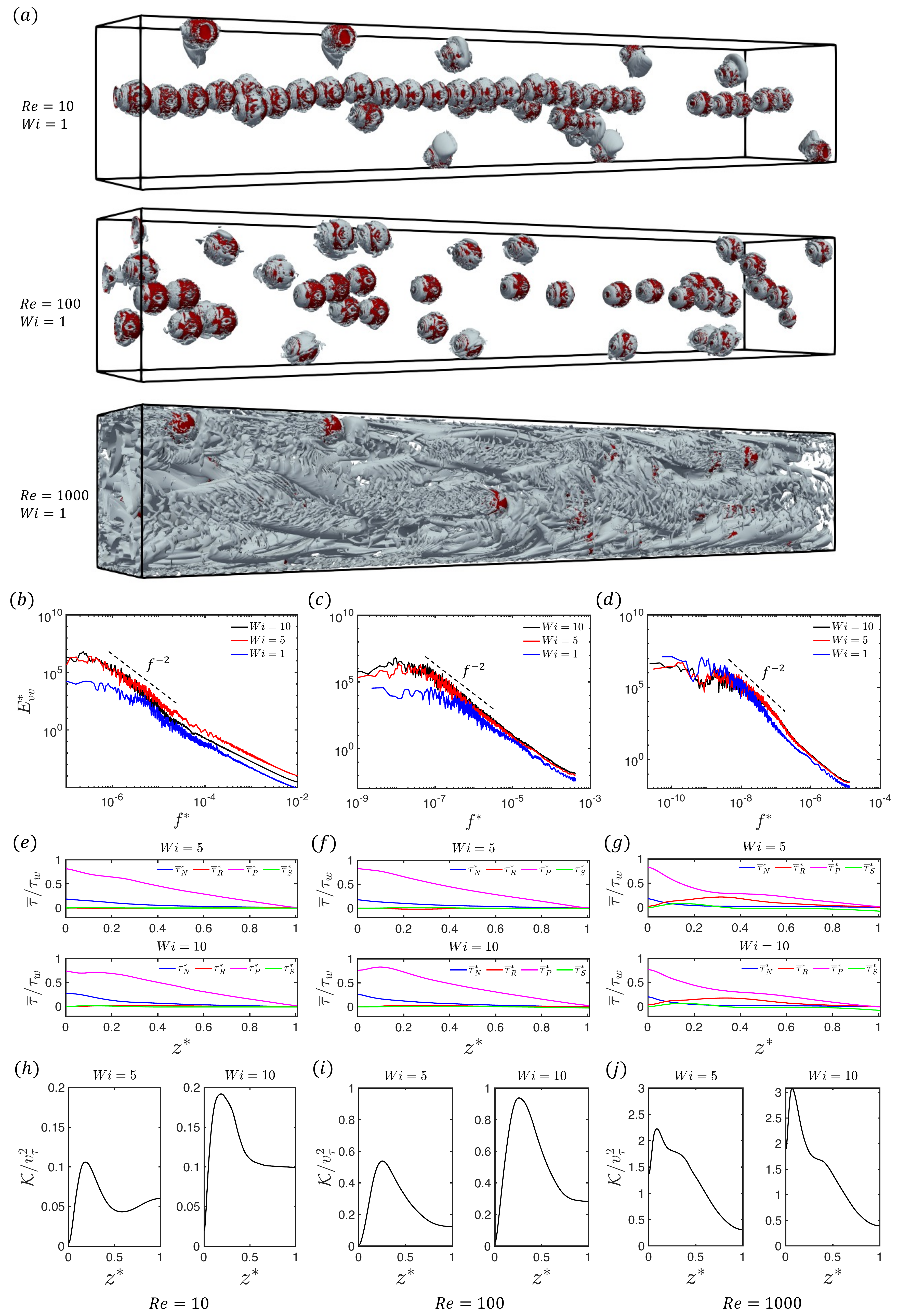}
\caption{\textit{(a)} Iso-surfaces of $Q-criterion$ are shown at different values of $Re$ for $Wi=1$. The contours are plotted at $0.003$, $0.3$ and $3$ for $Re=10, 100$ and $1000$ cases, respectively.  \textit{(b,c,d)} 1D energy spectrum, \textit{(e,f,g)} the distribution of various components of shear stress and \textit{(h,i,j)} turbulent kinetic energy ($\cal{K}$) are plotted in the mid-plane for $Re=10$ (\textit{left}), $Re=100$ (\textit{middle}) and $Re=1000$ (\textit{right}) cases, respectively ($\beta=0.1$ and $Ca=0.01$).}
\label{multiphase}
\end{figure}

As the flow instability is mainly triggered by the presence of the bubbles, the chaotic flow also remains restricted to the core region of the duct once the bubbles are aligned, as shown by the contours of $Q$-criterion in Fig.~\ref{multiphase}a. At this low values of $Re=10$ and $Wi=1$, even when the bubbles remain randomly distributed in a flow with high shear-thinning effect, the instability remains confined to immediate surroundings of the bubbles only. The velocity signals collected from the central point of the duct ($x^*=1, y^*=6, z^*=1$) for different values of $\alpha$ are shown in Fig.~\ref{alignment}b. The velocity shows random fluctuations in all three components when the bubbles are aligned close to the central point of the duct. Once the bubbles are randomly distributed, the velocity components show negligible fluctuations at the central point as shown in the inset of Fig.~\ref{alignment}b. When the inertia is increased ($Re=100$) at the same low value of $Wi=1$, majority of the bubbles accumulate in the core region in the form of small clusters. The flow still remains laminar at the walls due to low inertia and the presence of few bubbles there. Finally, once the inertia is very high ($Re=1000$), the bubbles become randomly distributed in the entire duct and, as a result, a fully chaotic flow regime is observed in the entire domain (Fig.~\ref{multiphase}a) with a significant increase in the friction drag.

Figure~\ref{multiphase}b shows 1D energy spectrum of the streamwise component of flow velocity for $Wi=1$, $Wi=5$ and $Wi=10$ at $Re=10$. Once the flow reaches a statistically steady state, the velocity signal is collected from the point away from the centre ($x^*=0.5, y^*=6, z^*=1$) to avoid the noise due to the bubble cluster. The flow velocity inside the bubbles is filtered out from the signal. A slope of $-2$ is observed for the $Wi=5$ \& $10$ cases when the flow is chaotic in the entire domain. This slope is the same as the temporal spectrum scaling of centreline velocity observed by \citet{lellep2024purely} indicating a similar chaotic state. The energy spectra of $Wi=5$ and $Wi=10$ cases overlap each other, showing a similar flow state for both the cases in the presence of bubbles. As the flow is not fully chaotic at $Wi=1$ and its signal shows a very high intermittency, the same is manifested in its spectrum where the slope does not match with the slopes of $Wi=5$ \& $10$ cases. For the $Re=100$ cases, the slope for $Wi=1$ gets closer to the slopes in the $Wi=5$ \& $10$ cases and finally at $Re=1000$, all the three cases show a similar slope of $-2$ (Fig.~\ref{multiphase}d). This slope of $-2$ is greater than the classical slope of $-5/3$ in the inertial turbulence but still less than $-4$ reported for a purely elastic turbulence in the absence of inertia (\citet{datta2022perspectives}) or at a very high value of $Wi=1000$ observed in the single-phase cases (Fig.~\ref{single-phase}a,c). As this multiphase EIT regime is achieved at the lower values of $Wi$, it is governed by both inertia and viscoelasticity, and hence the slope remains in between these two limits.

The distribution of different components of shear stress from the channel wall towards the centre are shown in Figs.~\ref{multiphase}e, f \& g in the mid-plane of the duct for $Re=10, 100$ \&
$1000$ cases, respectively. As the flow is multiphase, an additional contribution from the surface tension force of the bubbles ($\bar \tau_{\scaleto{S}{3.5 pt}}$) is added towards the total shear stress balance. However, this contribution remains much smaller than the remaining components as the bubble volume fraction is a mere $3\%$. Similar to the single-phase flow cases, these stress components are normalized by the local shear stress at the wall of the mid-plane. As seen in the figure, the chaotic flow regime is dominated by the viscoelastic stress ($\bar \tau_{\scaleto{P}{3.5 pt}}$) followed by the viscous stress due to the mean flow ($\bar \tau_{\scaleto{N
}{3.5 pt}}$). These two stress components are highest at the wall and decay to zero towards the channel centre where the shear rate is minimum. The contribution of the Reynolds stress ($\bar \tau_{\scaleto{R}{3.5 pt}}$) remains negligible for the $Re=10$ and $Re=100$ cases. The Reynolds stress component $\bar \tau_{\scaleto{R}{3.5 pt}}$ starts to show its effect once the inertia becomes high enough in the $Re=1000$ cases. As the source of flow instability is the curved streamlines around the bubble in this multiphase flow regime (\citet{mckinley1996rheological}), the transition to the EIT state greatly depends upon the distribution of bubbles across the channel. The role of viscoelasticity in promoting the bubble migration towards the channel centre or towards the wall is governed by the relative change in the convective time scale of the bubbles and the polymer relaxation time as has been explained by~\citet{bothe2022molecular}. A similar observation was also made in our earlier work as well (\citet{Naseer_Izbassarov_Ahmed_Muradoglu_2024}). The distribution of turbulent kinetic energy (TKE) for all these three values of $Re$ shows a peak value near the wall and minimum at the channel centre, as shown in Figs.~\ref{multiphase}h, i \& j. By increasing the value of $Wi$, the magnitude of TKE  also increases following the same qualitative trend in agreement to our previous study of viscoelastic turbulent single phase case (\citet{Izbassarov_Rosti_Brandt_Tammisola_2021}).

\begin{figure}
\centering
\includegraphics[width=0.5\textwidth]{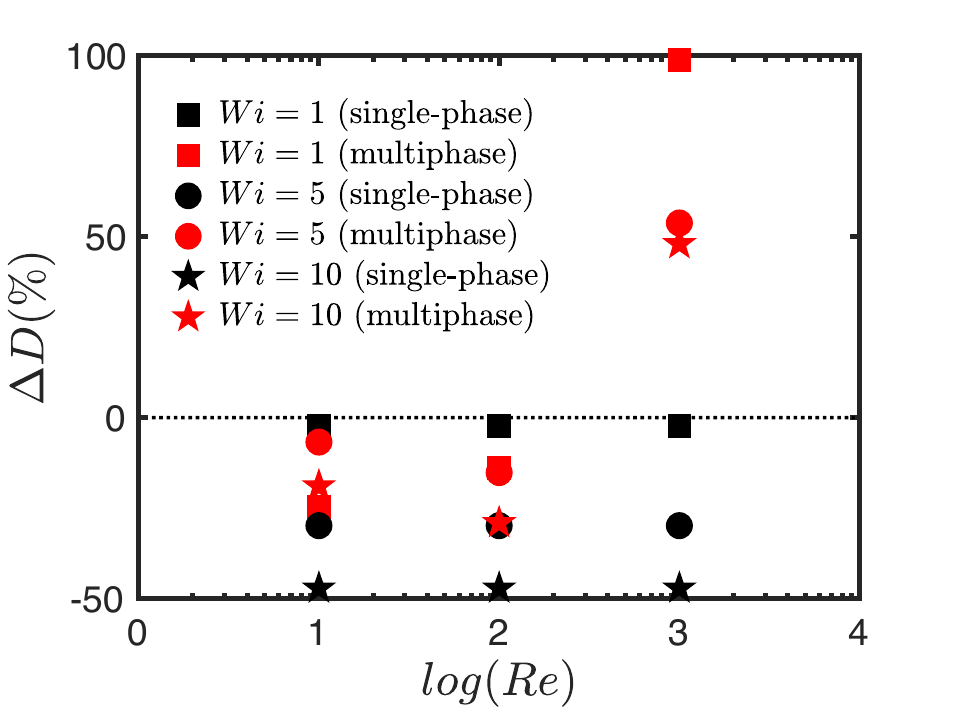}
\caption{The percentage change in drag measured by the change in total shear stress at the wall ($\overline \tau_w$) as compared to the Newtonian laminar flow once the flow is made viscoelastic and is subsequently transitioned to turbulent state by injecting the bubbles}
\label{DR}
\end{figure}
 
Once the flow reaches a steady state, the applied pressure gradient is balanced by the total shear stress at the walls and the average wall shear stress can be simply computed as $\overline \tau_w = -\frac{h}{2}\frac{dp_o}{dy}$. The change in the drag  is quantified by $\Delta D=(\overline \tau_{w} - \tau_{w_o})/\tau_{w_o}$, where $\tau_{w_o}$ is the wall shear stress of the corresponding Newtonian laminar flow for the same Reynolds number. Figure \ref{DR} summarizes the percentage change in drag for all the multiphase simulations. The single-phase viscoelastic laminar flow has lower drag as compared to the Newtonian laminar flow for the same value of $Re$. Understandably, as the inertia increases in this shear thinning viscoelastic fluid once the flow is made viscoelastic, the drag is reduced. For a particular value of $Wi$, the percentage reduction in drag is found to be the same for every value of $Re$. However, the drag increases significantly once a viscoelastic flow transitions to a chaotic turbulent state after injecting the bubbles (Fig.~\ref{DR}). In particular, a rapid increase in the drag is observed at $Re=1000$ once the flow becomes turbulent for all three values of $Wi$. It is interesting to observe that, although the drag increases for the $Re=10$ \& $100$ cases at $Wi=5$ \& $10$ upon transition to the EIT state, it still remains less than that of the respective Newtonian laminar state. Another interesting feature is also observed at $Wi=1$ for the $Re=10$ \& $100$ cases. Once bubbles are injected into the flow, minor fluctuations remain confined to the core region away from the walls where the bubbles collect and form a string-shaped pattern ($Re=10$) or small clusters ($Re=100$). The flow remains essentially laminar closer to walls and in the major portion of the duct. In this intermittent regime, the drag is reduced as compared to the single-phase laminar state. At a higher inertia ($Re=1000$), the flow becomes fully chaotic in the entire channel, and thus the drag is increased significantly even for the $Wi=1$ case (Fig.~\ref{DR}).
Furthermore, it is interesting to note that for $(Re,Wi)=(10,10)$ and $(100,10)$ cases, the flow has become completely chaotic and the EIT state is achieved by injecting the bubbles into the flow, however, the drag is still less than the corresponding Newtonian laminar state, which is primarily attributed to the shear-thinning effect of viscoelastic fluid (Fig.~\ref{DR}).
\subsubsection{TKE in a multiphase EIT regime}
\begin{figure}
\centering
\includegraphics[width=1.0\textwidth]{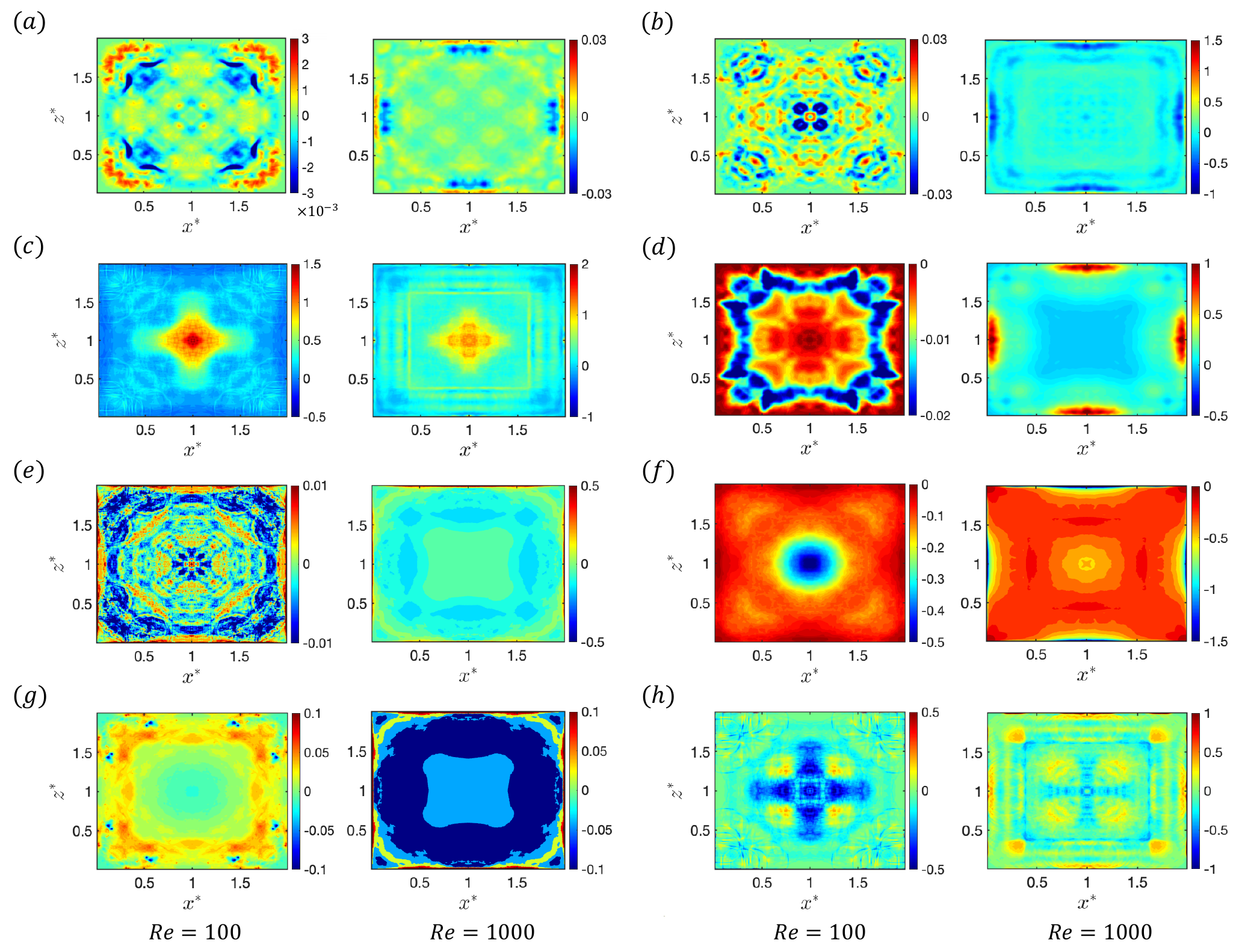}
  \caption{Contours of (\textit{a}) advection by mean flow, (\textit{b}) transport by velocity fluctuations, (\textit{c}) transport by pressure, (\textit{d}) production by mean flow, (\textit{e}) viscous diffusion, (\textit{f}) viscous dissipation, (\textit{g}) polymer work and (\textit{h}) body force terms are shown in a vertical cutting $xz$-plane for (\textit{left}) $Re=100, Wi=10$ and (\textit{right}) $Re=1000, Wi=10$ multiphase cases, respectively.}
\label{TKE_mp}
\end{figure}

To get further insight into the multiphase EIT regime, all the terms in Eq.~\ref{TKE} are integrated in the streamwise direction and averaged in time after the flow reaches a statistically steady state. The results are plotted in Fig.~\ref{TKE_mp} for $Re=100$ and $Re=1000$ at $Wi=10$. Compared to the single phase case shown in Fig.~\ref{TKE_sp}, a prominent effect of bubbles is observed in smoothing sharp gradients near the wall in all the terms that contribute to TKE. We note, however, that comparison between single and multiphase cases should be interpreted only qualitatively since $Wi$ is two orders of magnitude higher in the single phase case. Figure~\ref{TKE_mp} shows that the magnitudes of all terms are amplified as $Re$ is increased from 100 to 1000 but the amplification is more pronounced in advection (${\cal A}$), turbulent transport (${\cal Q}$),  production by mean flow (${\cal P}$) and viscous diffusion (${\cal D}$) terms. There are also some qualitative differences in these terms in the $Re=100$ and $Re=1000$ cases. The peak values of ${\cal A}$, ${\cal Q}$, $P$ and ${\cal \epsilon}$ occur in the corners and in the middle of the side walls for $Re=100$ and $Re=1000$, respectively. In contrast to the single phase case, the maximum value of ${\cal R}$ occurs in the central portion of the duct due to presence of bubbles (Fig.~\ref{TKE_mp}c). Significant viscous dissipation observed in the channel centre is also attributed to the presence of bubbles there. The negative peaks of polymer work for the multiphase case shows that it still contributes towards the production instead of dissipation in some portions of the domain (Fig.~\ref{TKE_mp}g). The body force term contributes significantly to the TKE budget now due to the work done by the surface tension (Fig.~\ref{TKE_mp}h).  
\section{Conclusions}\label{Conclusions}
Direct numerical simulations of single and multiphase viscoelastic fluid flows are performed in a square-shaped channel and the complex dynamics of elasto-inertial turbulent regime are explored without introducing any explicit perturbations from outside the system. The bubble interfaces are fully resolved using a finite-difference/front-tracking method coupled with the Giesekus model equations to simulate chaotic flow regime of a shear-thinning viscoelastic fluid. Instead of using conventional numerical schemes to discretize the constitutive equations by introducing artificial diffusion, a computationally expensive but robust log-conformation method is employed for the present DNS to preserve the integrity of EIT physics \citep{Yerasi_Picardo_Gupta_Vincenzi_2024}. The Reynolds number is varied by two orders of magnitude to cover a vast range below the threshold value of the inertial turbulence in the channel flow. Then the Weissenberg number is varied by three orders of magnitude at each of these values of Reynolds numbers to explore the conditions for triggering transition to a turbulent regime in this rectilinear channel flow. Subsequently, unlike any previously tested methods, bubbles are injected into the flow under the conditions for which a single-phase flow remains fully laminar to check whether the presence of bubbles can trigger an instability to achieve the EIT flow regime. A constant flow rate is maintained corresponding to each $Re$ by dynamically adjusting the applied pressure gradient. As different pathways can be followed in $Re$-$Wi$ space to achieve EIT (\citet{samanta2013elasto}), a concentrated polymer solution ($\beta=0.1$) is used in the present study for which the EIT regime is triggered much earlier than the inertial turbulence as the Reynolds number is increased.

It is observed that beyond $Wi \geq 100$, a single-phase flow becomes unstable when $Re \geq 100$. Instabilities start to grow in this straight channel even without introducing any discrete external perturbations. These instabilities grow with the passage of time, and ultimately a chaotic flow regime is observed in the entire domain. The threshold value of $Wi$ to trigger instability in the flow reduces with an increase in $Re$. For $Re=1000$, the critical value of $Wi$ to trigger transition from laminar to turbulent state is found to be $50$. When the value of $Re$ is low ($Re=10$), a Weissenberg number as high as $1000$ can not make the flow unstable. The energy spectra of this single-phase EIT regime reveal $-4$ scaling at the higher frequencies (dissipation range) while the classical $-5/3$ scaling is observed at the lower frequencies (inertial range) as the flow is dominated by both inertia as well as the elasticity. 

A novel mechanism of achieving the EIT state by injecting the bubbles into the flow under the conditions for which a single-phase flow remains laminar is tested and verified at a Reynolds number as low as $10$ and a Weissenberg number as low as $5$. The curved streamlines across the bubbles interfaces provide the necessary condition to trigger an instability even when the viscoelastic stresses are not very high and the single-phase flow remains completely laminar. The energy spectra shows a scaling of $-2$ for this low $Wi$ multiphase EIT regime. Thus, it is concluded that the scaling of energy spectra varies from $-4$ to $-5/3$ depending upon the relative dominance of inertia and elasticity, i.e, $-4$ for a purely elastic turbulence and $-5/3$ for a purely inertial turbulence. All the other values of energy scaling  would fall in between these two limits depending upon the relative strength of inertia ($Re$) and elasticity ($Wi$). It is also observed that the drag increases for all the situations where the laminar viscoelastic flow is fully transitioned to a turbulent state by injecting the bubbles into the flow. However, at the lower values of $Re=10$ \& $100$, this higher value of skin friction drag of the EIT regime still remains lower than that of the corresponding laminar Newtonian flow for the same Reynolds number. Once the inertia and the elasticity are both low, the bubbles are aligned in the core region of the duct forming a string-shaped pattern ($Re=10$, $Wi=1$) or clusters ($Re=100$, $Wi=1$), and the flow remains laminar in the wall region. For these cases, the friction drag reduces even further instead of increasing in this highly intermittent  flow. Unlike the solid particles, it is found that a higher shear-thinning effect breaks up the alignment of bubbles. 

The idea of achieving the EIT state at a very low value of $Re$ and $Wi$ just by injecting the bubbles into the flow can find many potential applications involving processes where mixing, heat transfer or other transport phenomena are of primary importance. The present study would intrigue experimental verification as it has been demonstrated that the requirement of $Wi$ to achieve the EIT state in a multiphase flow is low and realistic. 

\section{Data Availability}

Data can be shared upon request.
  
\section{Acknowledgment}

We acknowledge financial support from the Scientific and Technical Research Council of Türkiye (TUBITAK) [Grant Number 124M335] and the Research Council of Finland [Grant Number 354620].

\section{Declaration of Interest}

The authors declare no conflict of interest.

\begin{center}
\section*{Appendix}\label{app}
\end{center}

\begin{figure}
\centering
  \includegraphics[width=1.0\textwidth]{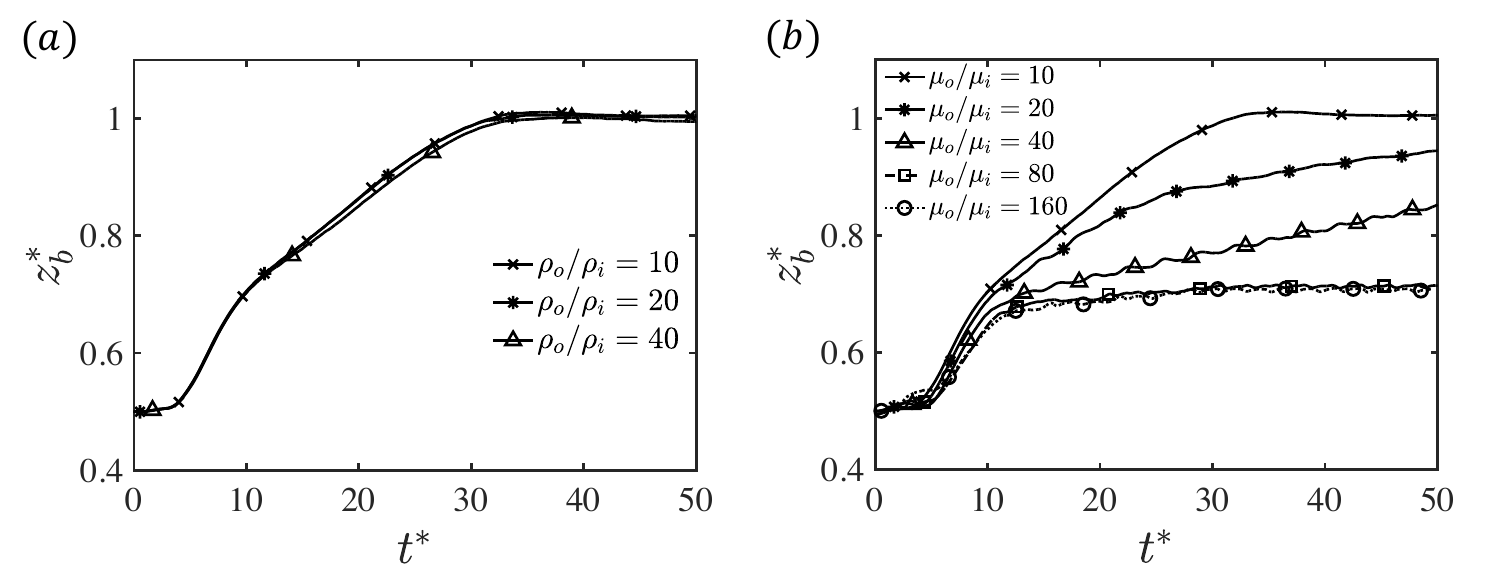}
  \caption{Effects of density and viscosity ratios. Evolution of a single bubble path in the wall-normal ($z$) direction in a concentrated polymer solution at (\textit{a}) different density and at (\textit{b}) different viscosity ratios ($Re=10, Wi=1, Ca=0.01, \beta=0.1$)}
\label{ratios}
\end{figure}

\noindent Further simulations are performed to examine the effect of density and viscosity ratios on the multiphase flows considered in this study. For this purpose, the migration of a single bubble is examined in the pressure-driven viscoelastic channel flow, and the density and viscosity ratios are varied in the ranges of $10 \leq \rho_0/\rho_i \leq 40$ and $10 \leq \mu_{o}/\mu_{i} \leq 160$. The other parameters are fixed at $Re=10, Wi=1, Ca=0.01$ and $\beta=0.1$. The results are shown in Fig.~\ref{ratios}. It is observed that the effects of density and viscosity ratios are negligible when $\rho_0/\rho_i \geq 10$ and  $\mu_{o}/\mu_{i} \geq 80$, as seen in Fig.~\ref{ratios}a and \ref{ratios}b, respectively. Therefore, the density and viscosity ratios are set to $\rho_0/\rho_i = 10$ and $\mu_{o}/\mu_{i} = 80$ in all the results presented in this paper. In a concentrated polymer solution (e.g., $\beta=0.1$) as considered in the present study, the viscosity ratio may have an impact on the bubble dynamics as the actual ratio can be as high as $10^8$. However, it is shown here that beyond $\mu_{o}/\mu_{i} \geq 80$, the difference in the bubble path becomes less than $1\%$ (Fig. \ref{ratios}).

\bibliographystyle{jfm}
\bibliography{References}
\end{document}